\documentclass[aps,prb,amsmath,amssymb,twocolumn,10pt,footinbib]{revtex4-1}
\usepackage{graphicx}
\usepackage{hyperref}

\begin{document}
\title{Metastable and spin-polarized states in electron systems with localized electron-electron interaction}

\author{Vladimir~A.~Sablikov and Bagun~S.~Shchamkhalova}

\affiliation{V.A. Kotel’nikov Institute of Radio Engineering and Electronics, Russian Academy of Sciences,
Fryazino, Moscow District, 141190, Russia}

\begin{abstract}
We study the formation of spontaneous spin polarization in inhomogeneous electron systems with pair interaction localized in a small region that is not separated by a barrier from surrounding gas of non-interacting electrons. Such a system is interesting as a minimal model of a quantum point contact, in which the electron-electron interaction is strong in a small constriction coupled to electron reservoirs without barriers. Based on the analysis of the grand potential within the self-consistent field approximation, we find that the formation of the polarized state strongly differs from the Bloch or Stoner transition in homogeneous interacting systems. The main difference is that a metastable state appears in the critical point in addition to the globally stable state, so that when the interaction parameter exceeds a critical value, two states coexist. One state has spin polarization and the other is unpolarized. Another feature is that the spin polarization increases continuously with the interaction parameter and has a square-root singularity in the critical point. We study the critical conditions and the grand potentials of the polarized and unpolarized states for one-dimensional and two-dimensional models in the case of extremely small size of the interaction region.
\end{abstract}
\maketitle

\section{Introduction}
\label{Intro}

Spin-related phenomena due to electron-electron (e-e) interaction in mesoscopic systems attract great interest, since they exhibit non-trivial physics of many-body systems and open new possibilities to manipulate the spin degrees of freedom. One of the most intriguing is the question of spontaneous breaking of the spin symmetry in quantum point contacts~\cite{Berggren_Pepper}. These structures are of interest also because they allow one to manipulate the spin and generate spin currents~\cite{Frolov,Koop}. However, currently there are fundamental physical problems in understanding their electronic structure and transport properties. Numerous experiments reveal transport features, such as puzzling $0.7\times 2e^2/h$ conductance anomaly observed at finite temperature, and other nonuniversal plateaus of the conductance arising at a finite voltage bias~\cite{Berggren_Pepper,Micolich}. Their nature remains a mystery whose solution lies in unclarified so far physics of interacting electrons in these systems.~\cite{Micolich2}. Nevertheless, it is clear that the origin of the anomalies is  closely related to the spin-charge structure of the quantum contact.

In the present paper we have found an unusual feature of the behavior of interacting electrons that could be a reason of the above anomalies. This feature arises when the phase transition with spontaneous breaking the spin symmetry occurs in the case where the e-e interaction is highly nonuniform, namely the interaction is concentrated in a small region of space that is not separated by any barrier from the surrounded gas of non-interacting electrons.

The problem is the following. The electrons in the quantum point contact are often considered as a one-dimensional (1D) system. However a rigorous theorem due to Lieb and Mattis~\cite{Lieb-Mattis} shows that the ground state of a 1D system is unmagnetized. In reality, the 1D part of the quantum point contact is continuously transformed at its ends  into surrounding system of higher dimensionality. Therefore this theorem is not applicable and the 1D segment can have a magnetic momentum. We draw attention to two facts. First, the e-e interaction in the narrow constriction is effectively much more strong than in the surrounding electron gas. Second, there is no physical reason to divide the system under consideration into a two coupled systems: a small system in which the e-e interaction is present, and a large system where electrons do not interact. The interaction region and the surrounding gas should be considered non-perturbatively as a single system. It is also important that under the equilibrium the charge and spin densities are formed in the constriction and the surrounding electron gas. Therefore, following questions arise: what spin and charge textures are formed in the equilibrium, whether the system can be spontaneously magnetized and under what conditions a magnetic momentum arises in the interaction region, what effective potential landscape is ultimately formed. 

In this paper we consider a minimal model that allows one to answer qualitatively these questions and to reveal non-trivial features of the spin-polarized state formation. The problem is solved within the self-consistent field approach by the way of minimizing the grand potential of the whole system. We come to an unexpected conclusion that the formation of a spin-polarized state strongly differs from Bloch or Stoner transition in homogeneous systems~\cite{Giuliani_Vignale,DasSarma}. It turns out that a metastable state appears in the critical point in addition to the globally stable state, and only one of these states is polarized.

The outline of the paper is the following. In Sec.~\ref{model} we present the model and the approaches used in the calculations of the grand potential and the electron densities. Sec.~\ref{1Dsystem} contains the analysis of a 1D model including the effect of an additional scatterer on the metastable state. In Sec.~\ref{2D}, a 2D system with the localized e-e interaction is considered. In Sec.~\ref{concluding} we discuss main results and possible applications. In Appendices we prove the existence of the branching in the critical point and analyze the stability of solutions.

\section{The model}\label{model}
\begin{figure}
\centerline{\includegraphics[width=.9\linewidth]{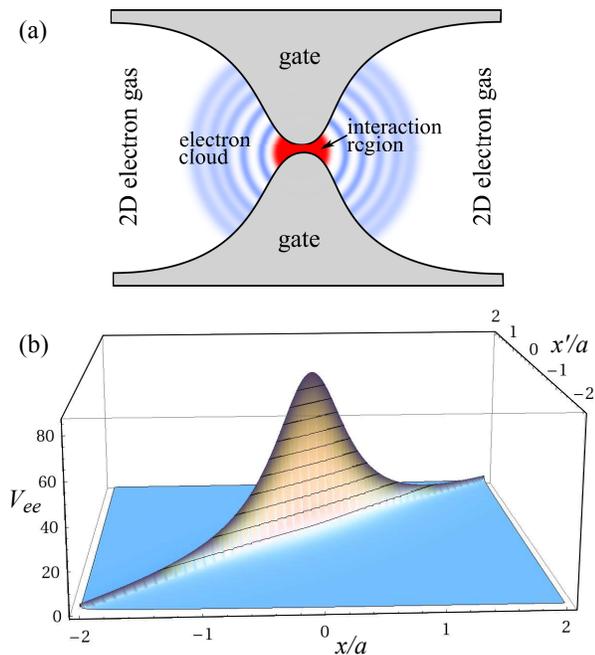}}
\caption{(Color online) (a) Schematics of quantum point contact. (b) The effective 1D interaction potential $V_{ee}(x,x')$ in the parabolic constriction $d(x)=d_0+x^2/a$, for $d_0/a=0.2$ and $n=n'=1$.}
\label{model_2}
\end{figure}

Consider 2D or 1D electron system in which the pair interaction potential $V_{ee}(\mathbf{r},\mathbf{r}=\mathbf{r}')$ is nonzero only in a finite region and vanishes outside it. Inhomogeneous interaction of this kind can actually be realized because of two reasons: (i) due to the screening of the Coulomb interaction by nearby conductors and  (ii) as a result of the confinement of electrons by lateral gates in 2D systems. The latter is realized in quantum point contacts, where the e-e interaction is effectively the strongest in the most narrow part of the constriction which is effectively one-dimensional. The effective interaction potential is estimated as~\cite{Shchamkhalova}
\begin{equation}
\left[V_{ee}(x,x')\right]_{ nn'}=\frac{e^2}{\epsilon}\iint\!\!dydy'\frac{\chi_{n,x}(y)|^2|\chi_{n',x'}(y')|^2}{\sqrt{|\mathbf{r}-\mathbf{r'}|^2+d_z^2}},                                                                                                                                \label{V_ee}
\end{equation}
where $\chi_{n,x}(y)$ is the transverse wave function of $n$-th subband, $d_z$ is the thickness of 2D layer, $\epsilon$ is the dielectric constant. The effective 1D pair interaction potential as a function of the longitudinal coordinates $x$ and $x'$ of interacting electrons is illustrated in Fig.~\ref{model_2}. 

Since the inhomogeneous interaction is caused by nearby gates, we include for generality into the model also a single-particle potential $U(\mathbf{r})$ created in the interaction region by external charges, such as a background charge and charges on the gates. 
 
The problem is to find the equilibrium densities of electrons with spin up and spin down in the interaction region and around it. To this end we will calculate the grand potential of the system and find the wave functions which minimize it. In this way there is a difficulty associated with taking into account the electron correlations which are strongest in the interaction region. The problem is simplified if one suppose that the size of the interaction region is small compared to the average distance between electrons. In this case we use the self-consistent field approximation without restrictions imposed by spin and spatial symmetry of the wave functions. Using this approximation for inhomogeneous systems has a decisive advantage since it is non-perturbative and goes far beyond the first-order expansion in the interaction~\cite{Fetter_Walecka}. Ultimately, this approach allows us to solve a highly nonlinear problem of self-consistent finding the wave functions with account of the charge and spin densities in the interaction region. 

Our study is based on the analysis of the grand potential with using the method developed by N.D.~Memrin~\cite{Mermin}. Calculations are carried out as follows. First, the self-consistent equations for the wave functions are obtained by minimizing the grand potential $\Omega$ over a restricted class of trial density matrices, which are chosen in the form of the equilibrium density matrix for non-interacting particles in an effective field. Stationary points of the grand potential yield self-consistent equations for the single-particle wave functions. Next, we show that these equations have several solutions and investigate their stability by analyzing the second variation of the grand potential. Finally, we compare the grand potentials of the stable solutions and study their dependence on the interaction strength and other parameters of the system.

The self-consistent equations for single-particle wave functions $\Psi_{\mathbf{k}s}(\mathbf{r})$ have the form of the Hartree-Fock equations in which the electron density matrix contains the Fermi distribution function. In what follows we consider a simplified case of short-range e-e interaction where
\begin{equation}
V_{ee}(\mathbf{r,r'})\approx v(\mathbf{r})\delta (\mathbf{r-r}').
\label{Vee-delta}
\end{equation}
In this case the self-consistent equations for single-particle wave functions are
\begin{equation}
\left[\frac{\mathbf{p}^2}{2m_e}\!+\!U(\mathbf{r})\!+\!v(\mathbf{r}) n_{\bar{s}}(\mathbf{r})\right]\Psi_{\mathbf{k}s}=\varepsilon_{\mathbf{k}s}\Psi_{\mathbf{k}s}\,,
\label{main_eq}
\end{equation}
where $\mathbf{k}$ is an orbital quantum number, $s$ is the spin, $\bar{s}=-s$, $n_{\bar{s}}$ is the electron density with the spin opposite to $s$,
\begin{equation}
n_{\bar{s}}(\mathbf{r})=\sum_{\mathbf{k'}}\!|\Psi_{\mathbf{k'}\bar{s}}(\mathbf{r})|^2 \rho^F_{\mathbf{k'}\bar{s}}\,,
\label{gen_ns}
\end{equation}
and $\rho^F_{\mathbf{k}s}=1/\left(1+\exp[\varepsilon_{\mathbf{k}s}/T]\right)$ is the Fermi distribution function.

The grand potential $\Omega$ in the stationary points, where $\delta\Omega=0$, is easily found in the form
\begin{equation}
\Omega=-T\sum_{\mathbf{k},s}\ln\left(\!1\!+\!\exp\left[\frac{\mu-\varepsilon_{\mathbf{k},s}}{T}\right]\right)-\int\!\! d\mathbf{r}v(\mathbf{r})n_{\uparrow}(\mathbf{r})n_{\downarrow}(\mathbf{r}).                                                                                                                                               \label{grand_pot_gen}
\end{equation}
Here, the last term is directly caused by the interaction energy. The first term also contains an interaction dependent part arising because of a change in the density of states due to the interaction. Physically this part originates from the electron cloud that is formed around the interaction region. The rest of the first term is the thermodynamic potential of the non-interacting electrons. Thus, the grand potential can be presented in the form:
\begin{equation}
\Omega=\Omega_{res}+\Omega_{int},
\end{equation}
where $\Omega_{res}$ is the grand potential of non-interacting electrons and $\Omega_{int}$ is the interaction-dependent part of $\Omega$, which is only important in what follows.

To separate $\Omega_{int}$ from $\Omega_{res}$ we proceed as follows. First, we consider a finite system with a discrete spectrum, and then make the transition to infinite system by the way of the asymptotic expansion of $\Omega$ in powers of the system size. In this expansion, the term, which changes proportionally to the volume of the system, is $\Omega_{res}$. The term, which is finite when the volume goes to infinity, represents $\Omega_{int}$.

\section{One-dimensional model\label{1Dsystem}}
Main features of the phase transition with the spin symmetry breaking in the systems with inhomogeneous interaction can be easily seen from the 1D model, which is simpler for calculations, and the results remain qualitatively in the 2D model. We assume that the size of the interaction region (as well as the size of the region where the built-in potential $U(x)$ is localized) is small compared with the electron wavelength, so that the wave function is nearly constant therein. Averaging over the interaction region in Eq.~(\ref{main_eq}) results in the following Hartree-Fock equation with an effective delta-like term:
\begin{equation}
\left[\frac{d^2}{dx^2}-2Q_{\bar s}\delta(x)+k^2\right]\psi_{ks}=0\,, 
\label{1D_HF}
\end{equation}
where the wave number $k=\sqrt{2m_e\varepsilon}/\hbar$,
\begin{equation}
Q_{s}=\frac{m_ea}{\hbar^2}\left[\bar u+\bar vn_s(0)\right]\,.
\label{Qs1D}
\end{equation}
Here $a$ is the size of the interaction region; $\bar u=\bar{U}-\bar{v}n_b$ is the average single-particle potential, which is the sum of the potential $\bar U$ induced by external gates and the potential $\bar{v}n_b$ originating from a background charge density $n_b$ in the interaction region; $\bar v$ is the average of the pair interaction amplitude defined in Eq.~(\ref{Vee-delta}); $n_s(0)$ is the electron density in the interaction region, which is determined by the occupied states with spin $s$:
\begin{equation}
n_s(0)=\sum_k|\psi_{ks}(0)|^2\rho^F_{k}\,.
\label{n_s(0)}
\end{equation}

Equation~(\ref{1D_HF}) is easy to solve. If the system has a finite length $L$, the wave function in the interaction region with zero boundary conditions at the ends reads as
\begin{equation}
\psi_{ks}(0)=\sqrt{\frac{2}{L}}\frac{k}{k+iQ_{\bar s}}\,.
\label{psi_1D}
\end{equation}
In the asymptotic limit $kL\to \infty$, the electron density in the interaction region is
\begin{equation}
n_s(0)\simeq\frac{2}{L}\int\limits_0^{\infty}dkD_s(k)\frac{k^2}{k^2+Q_{\bar s}^2}\rho^F_{k}\,,
\label{dens_1D}
\end{equation}
where $D_s(k)$ is the density of states
\begin{equation}
D_s(k)\simeq\frac{L}{2\pi}+\frac{1}{\pi}\frac{Q_{\bar s}}{k^2+Q_{\bar s}^2}\,. 
\label{DOS_1D}
\end{equation}

Combining Eqs.~(\ref{Qs1D}), (\ref{dens_1D}) and (\ref{DOS_1D}) we come to the following system of equations for $Q_s$:
\begin{equation}
Q_s=\frac{m_ea}{\hbar^2}\left[\bar u+\bar v\int\limits_0^{\infty}\frac{dk}{\pi}\frac{k^2}{k^2+Q_{\bar s}^2}\,\rho_k^F\right],                                                                                                                                                     \label{Qs_1D}
\end{equation}
where $s=\uparrow, \downarrow$. These equations determine the possible values ​​of $Q_s$, at which the grand potential has an extremum.

Our finding is that Eq.~(\ref{Qs_1D}) can have several solutions and this property is rather general for systems with spatially localized interaction. For each solution $Q_s$, one can find the corresponding electron density $n_{\bar s}(x)$ with the use of Eq.~(\ref{dens_1D}).

\subsection{Electron densities}
Let us find possible solutions of Eq.~(\ref{Qs_1D}) and their corresponding electron densities $n_s(x)$. It is convenient to use dimensionless quantities:
\begin{equation}
 y_s=\frac{Q_s}{k_F},\quad A=\frac{m_ea}{\hbar^2k_F}\bar u,\quad B=\frac{m_eL}{\pi \hbar^2}\bar v,\quad M=\frac{\mu}{T}\,.
\label{dimensionless}
\end{equation} 
Here $B$ is the interaction parameter, $A$ represents the built-in potential, $M$ represents the chemical potential and the temperature, $k_F$ is the Fermi wave vector. In these notations Eq.~(\ref{Qs_1D}) takes the form
\begin{equation}
 y_s=A+B\int\limits_0^{\infty}\frac{d\xi\,\xi^2}{\xi^2+y_{\bar s}^2}\,\rho(\xi),
\label{ys_1D}
\end{equation}
with $\rho(\xi)=1/(1+\exp[M(\xi^2-1)])$.

It is convenient to solve the system of Eqs~(\ref{ys_1D}) by reducing it to a single equation. Introduce an auxiliary function $F(y,B)$,
\begin{equation}
 F(y,B)=A+B\int\limits_0^{\infty}\frac{d\xi\,\xi^2}{\xi^2+y^2}\,\rho(\xi).
\label{F(yB)}
\end{equation} 
In terms of this function, Eq.~(\ref{ys_1D}) reads
\begin{equation}
 y=F(F(y,B),B).
\label{y(B)}
\end{equation} 

One of the roots of Eq.~(\ref{y(B)}) is easy to find. It coincides with the root of a more simple equation
\begin{equation}
y_0=F(y_0,B)\,,
\label{y0_B}
\end{equation}
as one can verify by the direct substitution of Eq.~(\ref{y0_B}) into Eq.~(\ref{y(B)}). Eq.~(\ref{y0_B}) has a single positive root $y_0=y_0(B)$, if $A+B>0$. This is easy to see by taking into account Eq.~(\ref{F(yB)}), which shows that the RHS of Eq.~(\ref{y0_B}) decreases monotonically with $y_0$ and equals $A+B$ at $y_0=0$. The LHS of Eq.~(\ref{y0_B}) increases as $y_0$. Hence, Eq.~(\ref{y0_B}) has a single root. This root corresponds to an unpolarized state of the system. Indeed, in this case $y_s=y_{\bar s}=y_0(B)$. Thus, the system has the unpolarized state which exists at $A+B>0$ for any interaction parameter.~\cite{note1}

We have found that Eq.~(\ref{y(B)}) has another two roots in addition to $y_0(B)$. They arise when the interaction parameter exceeds a critical value $B_c$. The proof of this statement is given in Appendix~\ref{appendix_branching_1D} where we show that the function $y(B)$ defined by Eq.~(\ref{y(B)}) has a branching point at $B=B_c$, in which two roots, $y_{s}(B)$ and $y_{\bar s}(B)$, arise in addition to $y_0(B)$, and study their dependence on $B$. The additional roots correspond to the polarized state, since $y_{s}\neq y_{\bar s}$ and consequently the electron densities with opposite spins are different. In the vicinity of the branching point, $y_s(B)$ varies as $y_s-y_0\propto s(B-B_c)^{1/2}$, with $s=\pm 1$.

\begin{figure}
\centerline{\includegraphics[width=0.85\linewidth]{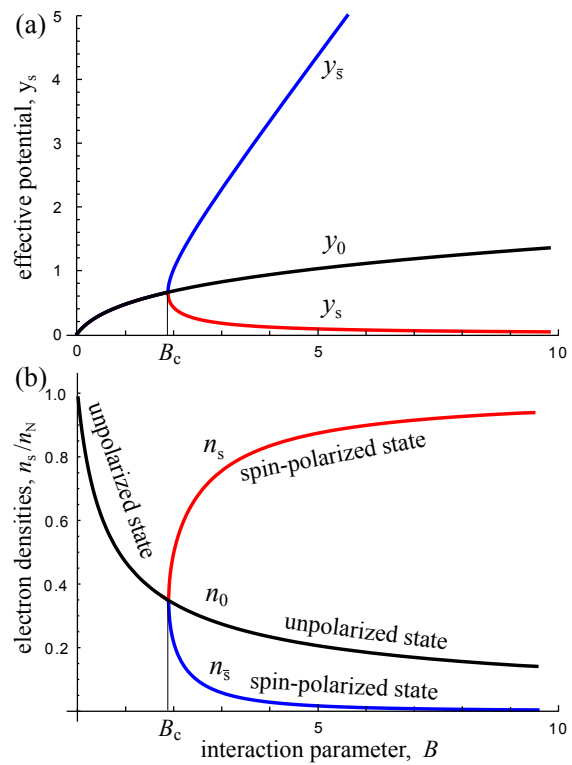}}
\caption{(Color online) Dimensionless effective potentials of the interaction region (a) and the electron densities in the interaction region (b) as functions of the interaction parameter for $T=0$ and $A=0$.}
\label{fig_yn(B)_1D}
\end{figure}

In the case of zero temperature, Eq.~(\ref{y(B)}) is explicitly solved in a wide range of $B$ since the integral in Eq.~(\ref{F(yB)}) is calculated analytically and the resulting transcendental equation is easily solved numerically. The results are illustrated in Fig.~\ref{fig_yn(B)_1D} where the roots of Eq.~(\ref{y(B)}) and the electron densities in the interaction region are shown as functions of the interaction parameter. Here and henceforth the electron density is normalized by $n_N=k_F/\pi$. It is seen that at the critical point both graphs, $y_s(B)$ and $n_s(B)$, split into polarized and unpolarized branches which coexist if $B>B_c$. Thus, above the critical point the system can be in two states: the unpolarized state with $n_{\uparrow}=n_{\downarrow}=n_0$ and the polarized state with $n_{\uparrow}\ne n_{\downarrow}$. The effective potentials of the interaction region in these states are different and depend on the spin. The electrons with spin up feel the potential produced by the spin-down electrons and vice-versa. Near the critical point, when $B-B_c\ll B_c$, the densities $n_{\uparrow}$ and $n_{\downarrow}$ deviate symmetrically from the unpolarized branch $n_0$, as Eq.~(\ref{A_ys_B}) shows, so that $n_{\uparrow}+n_{\downarrow}\simeq 2n_0$. However, far from the critical point this symmetry is violated because of the nonlinear dependence of the effective potentials on the interaction strength.

\begin{figure}
\centerline{\includegraphics[width=0.9\linewidth]{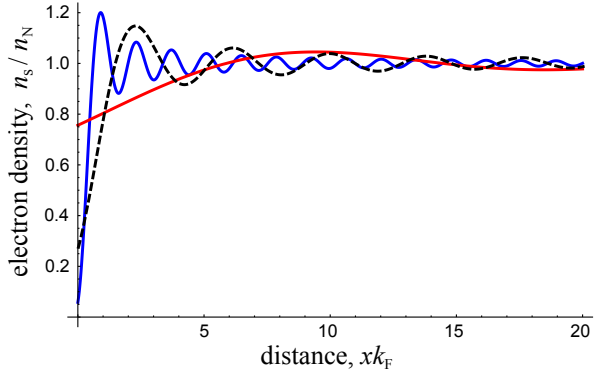}}
\caption{(Color online) The spatial dependence of the electron densities with opposite spin outside the interaction region for polarized (full lines) and unpolarized (dashed line) states. The parameters used in the calculations are $T=0$, $B=3$ and $A=0$.}
\label{fig_cloud_dens}
\end{figure}

In the surrounding electron gas, the electron density is also perturbed due to the interaction, and a spin-polarized electron cloud is formed. The electron densities with opposite spin outside the interaction region read:
\begin{equation}
n_s(x)=\!\int\limits_0^{\infty}\!\frac{dk}{\pi}\left[1\!-\!\frac{Q_{\bar s}}{\sqrt{k^2+Q_{\bar s}^2}}\cos\!\left(\!2kx\!+\!\arctan\frac{k}{Q_{\bar s}}\!\right)\right]\rho_k^F.
\end{equation} 
The spatial dependence of $n_s(x)$ is illustrated in Fig.~\ref{fig_cloud_dens} for both polarized and unpolarized states. In the polarized state, the densities of electrons with opposite spin oscillate with different periods and amplitudes while in the unpolarized state usual Friedel oscillations are formed.

Above calculations were carried out supposing that $A+B>0$. In reality $A+B$ can be negative when single-particle potential $\bar u=\bar{U}-\bar{v}n_b$ is negative. In this case a bound state arises for that spin component $s$, which feels the attracting effective potential ($Q_{\bar s}<0$). It is easy to generalize the results by adding the bound state with the wave function
\begin{equation}
 \psi^{(bs)}_{s}=\sqrt{-Q_{\bar s}}\left[e^{Q_{\bar s}x}\theta(x)+e^{-Q_{\bar s}x}\theta(-x)\right]\theta(-Q_{\bar s}).
\end{equation} 
However, direct calculations show that the presence of the bound state does not qualitatively change the dependence of $n_s$ on $B$. 

\begin{figure}
\centerline{\includegraphics[width=0.9\linewidth]{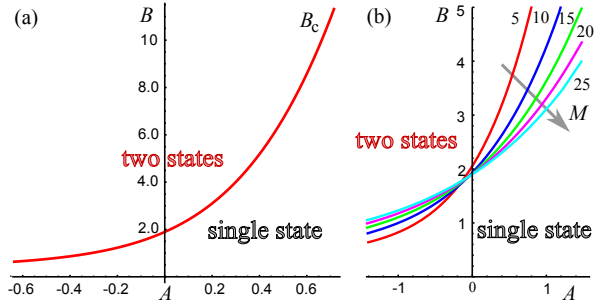}}
\caption{(Color online) Phase diagram of single (unpolarized) and two (unpolarized+polarized) states in the plane $(A,B)$. The lines image $B_c$ as a function of $A$. (a) The case of zero temperature and $M=\infty$. (b) The case of finite temperature. The variety of lines $B_c(A)$ corresponds to different values of $M=\mu/T$, which are specified near the lines.}
\label{fig_B(A)_T}
\end{figure}

The critical value of the interaction parameter $B_c$ depends on the parameters of the system: $A$ and $M$. The dependence of $B_c$ on $A$ is interesting because it simulates the effect of the gate voltage in the experiments on the quantum point contacts. The dependence of $B_c$ on $A$ is shown in Fig.~\ref{fig_B(A)_T}a for zero temperature. $B_c$ is seen to increase with $A$. The line $B_c(A)$ divides the plane ($A,B$) into two regions. Below this line, there is only the unpolarized state. Above the line, both polarized and unpolarized states exist. 

The effect of temperature was studied in the case where $T\ll \varepsilon_F$ (see Fig.~\ref{fig_B(A)_T}b). On a qualitative level, the temperature effect is the following. With increasing the temperature, $B_c$ strongly increases for positive $A$ and slightly decreases when $A$ is negative. When $A$ is close to zero, the variation of $B_c$ is more complicated, but generally $B_c$ slightly increases with the temperature. It is interesting that the critical point exists at any values ​​of the parameters for which the model we used is physically reasonable.

In the case where two states are available to the system, one needs to compare their grand potentials.

\subsection{Grand potential}
The grand potential of each state can be found from the general Eq.~(\ref{grand_pot_gen}). In the case of 1D model with small interaction region, the grand potential has the form
\begin{equation}
\begin{split}
 \Omega=&-\frac{TL}{2\pi}\int\limits_0^{\infty}dk\ln\left(1+e^{\frac{\mu-\varepsilon_k}{T}}\right)\\
 &-\frac{T}{\pi} \sum_sQ_s\int\limits_0^{\infty}\frac{dk}{k^2+Q_s^2}\ln\left(1+e^{\frac{\mu-\varepsilon_k}{T}}\right)\\ &-\bar{v}an_{\uparrow}(0) n_{\downarrow}(0).
\end{split}
\end{equation} 
Here the first term proportional to $L$ is $\Omega_{res}$, the second and third terms are $\Omega_{int}$. By excluding $n_s(0)$ with the use of Eq.~(\ref{Qs1D}) we get $\Omega_{int}$ in terms of $y_s$: 
\begin{equation}
\begin{split}
 \frac{\pi}{\mu}\Omega_{int}=&-\frac{1}{M}\sum_sy_s\int\limits_0^{\infty}\frac{d\xi}{\xi^2+y_s^2}\ln\!\left[1+e^{M(1-\xi^2)}\right]\\ &-\frac{2}{B}(y_{\uparrow}-A)(y_{\downarrow}-A).
\end{split}
\label{Omega_1D}
\end{equation} 

\begin{figure}
\centerline{\includegraphics[width=0.85\linewidth]{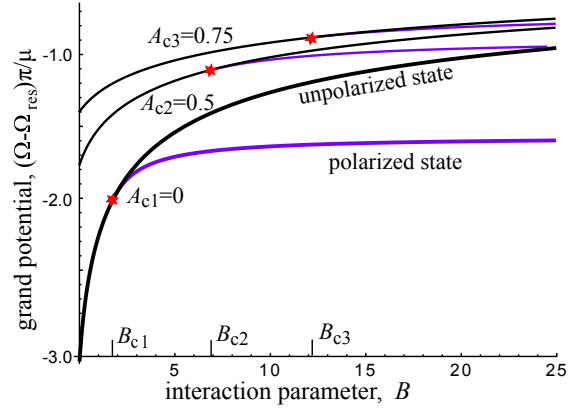}}
\caption{(Color online) Grand potential as a function of the interaction parameter for a variety of $A$ at $T=0$. Asterisks indicate the critical points. Black lines represent the unpolarized state, violet (gray) lines correspond to the polarized state.}
\label{Omega_B_1}
\end{figure}

Direct calculation of $\Omega_{int}$ with using the roots of Eq.~(\ref{y(B)}) leads to the results shown in Fig.~\ref{Omega_B_1}.

The grand potential of the polarized state $\Omega_P$ is seen to be lower than the grand potential of the unpolarized state $\Omega_0$ for all $A$. Near the critical point the difference of the grand potentials of the polarized and unpolarized states varies with $B$ as follows: $\Delta\Omega\equiv\Omega_P-\Omega_0\propto -(B-B_c)^2$.

Consider now the stability of the solutions. The analysis of the second variation of $\Omega$ is carried out in Appendix~\ref{appendix_stability} by expanding $\Omega$ up to second order in the variation of the self-consistent field for a finite system and then making the transition to the limit $L\to \infty$. We show that in this limit the matrix $\delta^2\Omega$ is positive for both (polarized and unpolarized) states and therefore the stationary points are local minima. The state with lower $\Omega$ corresponds to a global minimum and the other state therefore is metastable.

\subsection{Effect of a scatterer\label{scatterer}}
The above calculations show that the grand potential of the polarized state is lower than that of the unpolarized state. An interesting question is whether this is a general property, or the grand potential of the polarized state can be higher than that of the unpolarized state? That would be interesting, since in this case one can expect an unusual temperature dependence of the polarization. Unfortunately we failed to come to an universal conclusion about the sign of the grand potential difference $\Delta\Omega$ in general case, but in this section we show that $\Delta\Omega$ can really be positive.

The important point is that the difference between the grand potentials of the polarized and unpolarized states can be changed due to factors affecting the energy of the electron cloud around the interaction region. We study the effect produced by an additional scatterer located at some distance from the interaction region in the non-interacting electron gas. It turns out that in this case $\Delta\Omega$ changes dramatically.

Consider a 1D system containing a $\delta$-like center located at a distance $l$ from the interaction region. In this case the Hartree-Fock equation differs from Eq.~(\ref{1D_HF}) by the additional $\delta$-like potential
\begin{equation}
\left[\frac{d^2}{dx^2}-2Q_{\bar s}\delta(x)-2u_1\delta(x-l)+k^2\right]\psi_{ks}=0,
\label{HF_eq_scatterer}
\end{equation}
where $Q_s$ is defined by Eq.~(\ref{Qs1D}), $u_1$ is the amplitude of the scatterer potential. The problem is solved straightforwardly. After cumbersome calculations we arrive at the following results.

The phase transition with the formation of the metastable state persists in the presence of the $\delta$-like scatterer. The system is unpolarized when the interaction parameter $B$ (defined as before by Eq.~(\ref{dimensionless})) is less than a critical value $B_c$, which however depends on the amplitude $u_1$ of the scatterer potential and its position.  When $B>B_c$, a polarized state arises in addition to the unpolarized state. The dependence of the electron densities $n_s$ and $n_{\bar s}$ in the interaction region on $B$ is qualitatively similar to that in the case of the system without the scatterer.

\begin{figure}
\centerline{\includegraphics[width=.9\linewidth]{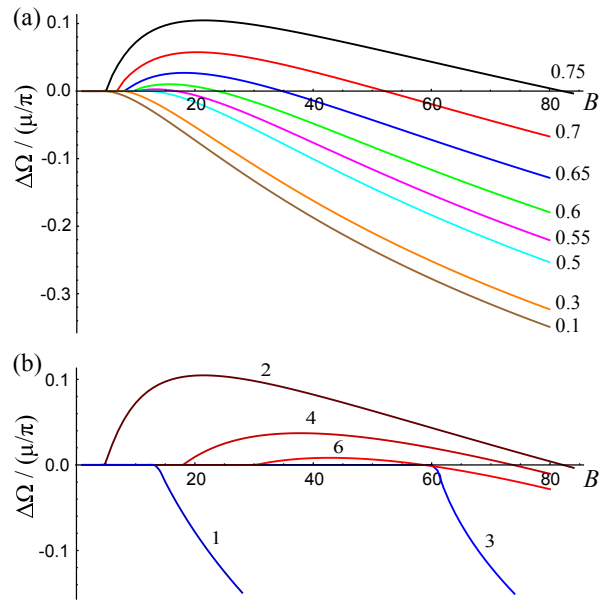}}
\caption{(Color online) (a) The difference in the grand potentials $\Delta\Omega$ of the polarized and unpolarized states as a function of the interaction parameter for a variety of the scatterer potential amplitude $u_1/k_F$ from 0.1 to 0.75 at $l=2\pi/k_F$. (b) $\Delta\Omega$ versus $B$ for a variety of the distance $l=2\pi n/k_F$, where $n$=1, 2, 3, 4, 6, at fixed $u_1/k_F$=0.75.}
\label{fig_grand_pot_scatterer}
\end{figure}

The difference between the grand potentials of the polarized and unpolarized states essentially depends on the position of the scatterer and its potential. The effect of the scatterer potential $u_1$ is different depending on the distance $l$. If the distance is close to an integer number of the Fermi wavelengths, the increase of $u_1$ results in the growth of the grand potential of the polarized state, so that $\Delta\Omega$ becomes positive in some range of $B$ near and above the critical point. This means that the polarized state becomes metastable in some range of $B$. However, a further increase of $B$ results in that the of grand potential difference becomes negative again. The specific results obtained for $lk_F=2\pi$ are presented in Fig.~\ref{fig_grand_pot_scatterer}a.

If the distance $l$ is close to half-integer number of the Fermi wavelengths, the critical value of the interaction parameter $B_c$ increases with the scatterer potential $u_1$. Specific results for the dependence of the grand potential difference on the interaction parameter are given in Fig.~\ref{fig_grand_pot_scatterer}b for a variety of $l$. 

\section{Two-dimensional system\label{2D}}

The question of whether the metastable state is formed in 2D systems is interesting because the configuration of the electron cloud in this case strongly differs from that considered above. This is moreover interesting, since in the realistic case of quantum contact the electron cloud is formed in 2D electron gas.

Consider a 2D electron system in which the e-e interaction is localized in a circle of radius $a$. In this case the Hartree-Fock equation~(\ref{main_eq}) reads
\begin{equation}
 \nabla^2\Psi_{\mathbf{k}s}+[k^2-\mathcal{Q}_{\bar s}^2(\mathbf{r})]\Psi_{\mathbf{k}s}=0,
\label{2D-HF}
\end{equation} 
where 
\begin{equation}
 \mathcal{Q}_{\bar s}^2(\mathbf{r})=\frac{2m_e}{\hbar^2}[v(r)n_s(\mathbf{r})+U(r)].
\label{2D-Qs}
\end{equation}

Let us assume again that the radius $a$ is small compared to the wavelength, $ak\ll1$, and treat the interaction region as a boundary condition for the wave function in the outer region. By integrating Eq.~(\ref{2D-HF}) over the interaction region, one obtains the following boundary condition:
\begin{equation}
 a\left.\frac{\partial\Psi_{\mathbf{k}s}}{\partial r}\right|_{r=a+0}+\frac{a^2}{2}Q_{\bar s}^2\Psi_{\mathbf{k}s}(r=a)=0,
\label{2D-bound_cond}
\end{equation} 
where $Q_s^2$ is the average of $\mathcal{Q}_s^2$ over the interaction region
\begin{equation}
 Q_s^2=\frac{1}{\pi a^2}\int\limits_0^adr\,r\int\limits_0^{2\pi}d\varphi\mathcal{Q}_s^2(\mathbf{r}).
\label{2D_Qs-average}
\end{equation} 

In the outer region the wave function reads as
\begin{equation}
 \Psi_{\mathbf{k}s}=\frac{1}{2\pi}e^{im\varphi}\psi_{kms}(r)\chi_s,
\label{2D_Psi}
\end{equation} 
where $\chi_s$ is the spin function, $m$ is integer, $\varphi$ is the angular coordinate,
\begin{equation}
 \psi_{kms}(r)=A_{kms}[J_m(kr)+B_{kms}Y_m(kr)],
\label{2D_psi}
\end{equation} 
where $J_m(kr)$ and $Y_m(kr)$ are the Bessel functions of the first and second kind. The coefficient $B_{kms}$ is determined from 
Eq.~(\ref{2D-bound_cond}). In what follows, only the coefficient $B_{kms}$ at $m=0$ is important, since all components of the wave function with $m\ne0$ are small in the interaction region. They are of the order of $(ak)^m$. For $B_{k0s}$ we have
\begin{equation}
 B_{k0s}\equiv B_{ks}=-\frac{2akJ_0'(ak)+a^2(k^2-Q_{\bar s}^2)J_0(ak)}{2akY_0'(ak)+a^2(k^2-Q_{\bar s}^2)Y_0(ak)},
\end{equation} 
here and below the index $m=0$ is dropped. Taking into account that $ak\ll1$, this equation can be simplified to
\begin{equation}
 B_{ks}=-\frac{\pi}{2}\frac{a^2Q_{\bar s}^2}{2-a^2Q_{\bar s}^2(\gamma+\ln(ak/2)},
\label{2D_Bks}
\end{equation} 
where $\gamma$ is Euler's constant.

The normalization constant $A_{ks}$ is found assuming that the system is infinite:
\begin{equation}
 |A_{ks}|^2=\frac{1}{1+|B_{ks}|^2}.
\label{2D_Aks}
\end{equation} 

The electron densities $n_{\uparrow}(0)$ and $n_{\downarrow}(0)$ in the interaction region are calculated using Eqs~(\ref{gen_ns}), (\ref{2D_Psi}), (\ref{2D_psi}), (\ref{2D_Bks}) and (\ref{2D_Aks}),
\begin{equation}
 n_s(0)=\frac{1}{2\pi}\int\limits_0^{\infty}\frac{dk\,k\,\rho_k^F}{\left[1\!-\!\frac{a^2Q_{\bar s}^2}{2}\left(\gamma\!+\!\ln\frac{ak}{2}\right)\right]^2\!+\!\frac{\pi^2}{4}\frac{a^4Q_{\bar s}^4}{4}}\,.
\label{2D_ns0}
\end{equation} 

Now one can obtain a system of self-consistent equations for the effective potentials of the interaction region $Q_s$. Using Eqs~(\ref{2D-Qs}), (\ref{2D_Qs-average}) and (\ref{2D_ns0}) we get
\begin{equation}
 \frac{Q_s}{k_F^2}=A+\frac{B_{2D}}{k_F^2}\int\limits_0^{\infty}\frac{dk\,k\,\rho_k^F}{\left[1\!-\!\frac{a^2Q_{\bar s}^2}{2}\left(\gamma\!+\!\ln\frac{ak}{2}\right)\right]^2\!+\!\frac{\pi^2}{4}\frac{a^4Q_{\bar s}^4}{4}},
\label{2D_equation-Qs}
\end{equation} 
where dimensionless parameters are introduced:
\begin{equation}
 A_{2D}=\frac{\bar{U}}{\mu}\,,\quad B_{2D}=\frac{m_e\bar{v}}{2\pi\hbar^2},
\end{equation} 
$\bar{U}$ and $\bar{v}$ denote $U(r)$ and $v(r)$ averaged over the interaction region. The value $\bar{v}$ can be roughly estimated as the average of the pair interaction potential $V_{ee}$ using Eq.~(\ref{Vee-delta}): $\bar{v}\sim \pi a^2\bar{V}_{ee}$.

The equation system~(\ref{2D_equation-Qs}) for $Q_s$ and $Q_{\bar s}$ is easily reduced to the single equation which has the same form as Eq.~(\ref{y(B)}), but now the function $F(y,B)$ is more complicated:
\begin{equation}
\begin{split}
 F(y_s,&B_{2D})=A_{2D}\\
&+B_{2D}\!\int\limits_0^{\infty}\frac{d\xi\,\xi\,\rho(\xi)}{\left[1\!-\!2\alpha^2y_s\left(\gamma\!+\!\ln(\alpha \xi)\right)\right]^2\!+\!\frac{\pi^2}{4}\alpha^4y_s^2}\,,
\end{split}
\label{2D_F(yB)}
\end{equation} 
where the dimensionless variables are used:
\begin{equation}
 y_s=\frac{Q_s^2}{k_F^2}\,,\quad \xi=\frac{k}{k_F}\,,\quad \alpha=\frac{ak_F}{2}\,.
\end{equation} 

Eq.~(\ref{y(B)}) with the function $F(y,B_{2D})$ defined by Eq.~(\ref{2D_F(yB)}) admits an analytical analysis similar to that given in Appendix~\ref{appendix_branching_1D}. As a result of analytical and numerical calculations we have found that in the 2D case the branching of solutions occurs similar to that described in Sec.~\ref{1Dsystem}. If the interaction parameter $B_{2D}$ is less than the critical value $B_{2D,c}$, the system is unpolarized. When $B_{2D}>B_{2D,c}$ there are two states one of which is polarized and the other unpolarized. Near the critical point, the effective potentials $y_s$ and the electron densities $n_s(0)$ change with $B\!-\!B_c$ similar to the above 1D model. 

The grand potential is expressed in terms of $y_s$ with using Eqs~(\ref{grand_pot_gen}), (\ref{2D_ns0}) and (\ref{2D_equation-Qs}). The interaction dependent part of the grand potential reads
\begin{equation}
\begin{split}
 \frac{\Omega_{int}}{\mu}&=\frac{2}{M}\sum_s\int\limits_0^{\infty}\frac{d\xi}{\xi}\frac{\ln\left[1+e^{M(1-\xi^2)}\right]}{\pi^2+4\left[\frac{1}{2\alpha^2y_s}-\gamma-\ln(\alpha \xi)\right]^2}\\
&-\frac{\alpha^2}{B_{2D}}(y_{\uparrow}-A_{2D})(y_{\downarrow}-A_{2D})\,.
\end{split}
\label{2D_grand-pot}
\end{equation} 

\begin{figure}
\centerline{\includegraphics[width=0.9\linewidth]{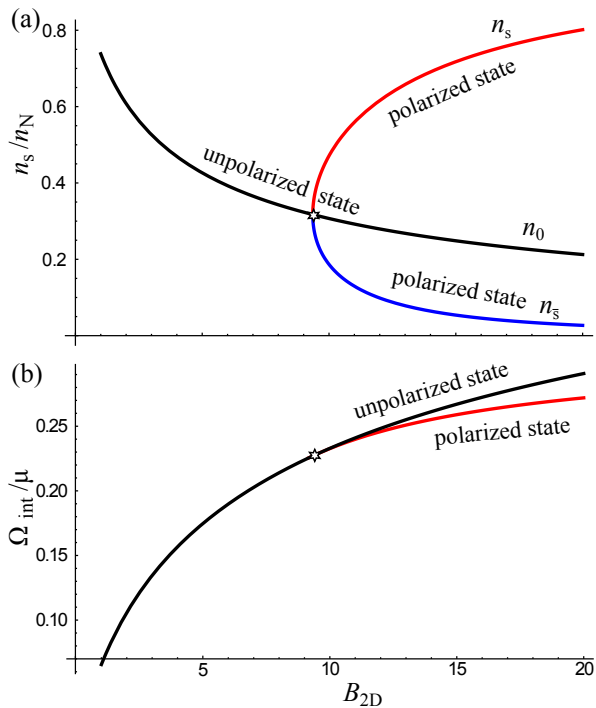}}
\caption{(Color online) (a) Electron densities in the interaction region and (b) the grand potential of the unpolarized and polarized states as functions of the interaction parameter. Calculations were carried out for $T=0$, $A_{2D}=0$ and $\alpha=0.3$.}
\label{fig_2D_n-Omega}
\end{figure}

The dependence of the electron densities in the interaction region and the grand potential on the interaction parameter is illustrated in Fig.~\ref{fig_2D_n-Omega}. It is seen that both $n_s(0)$ and $\Omega_{int}$ vary with $B$ quite similar to the 1D model.

\section{Discussion and concluding remarks\label{concluding}}

We have found a non-trivial behavior of interacting electrons in the inhomogeneous system where the e-e interaction is localized in a finite region, which is not separated by any barriers from surrounding non-interacting electron gas. Under this condition, no states are localized in this region, unlike the Kondo systems or quantum dots connected to electronic reservoirs. It turns out that in such a system, the phase transition with spontaneous breaking the spin symmetry due to exchange interaction is very different from the Bloch or Stoner transition in homogeneous systems.

The main feature is that a metastable state emerges at the critical point in addition to the globally stable state, so that both states exist above the critical point and only one of them is spin polarized. In other words, the metastable state can be either polarized or unpolarized. This depends on the form of the interaction region, its potential and the spatial configuration of the electron cloud around the interaction region. Another feature is that above the critical point, the spin polarization increases continuously with increasing the interaction parameter in contrast to the Bloch transition, where the polarization rises abruptly. These properties do not qualitatively depend on the dimensionality (they are similar for 1D and 2D systems) and persist in the presence of additional scatterers.

An important factor determining the formation of the metastable state is that the inhomogeneous system is not locally neutral. Excess charge and spin densities arise in the interaction region, so that the spin-dependent potential is formed self-consistently. This contrasts to homogeneous systems which are naturally supposed to be locally charge neutral~\cite{Giuliani_Vignale}. The spin and charge densities arising in the interaction region play a key role in our consideration. Their effect has been taken into account within the self-consistent field approach without restrictions imposed by the symmetry of wave functions. This approach is in line with recent studies of the spontaneously broken symmetry states of interacting electrons within unrestricted Hartree-Fock approximation which reveal spin and charge structure of the correlated state~\cite{Zhang,Baguet}. In our case this approach has advantage of being non-perturbative, but it loses the effect of dynamic correlations, which requires further study.

The metastable state can be realized in inhomogeneous electron systems made on the basis of 2D gas with the use of gates, such as quantum point contacts. Simple estimates show that the critical condition for the metastable state to appear is really attainable in such structures. The potential of pair interaction is approximated as: $V_{ee}\approx e^2/(\epsilon \sqrt{(x-x')^2+d^2})$, with $d$ being the width of the constriction. In the constriction of length $a$, the average of the interaction amplitude can be estimated as: $\bar{v}\approx (2e^2/\epsilon)\ln(2a/d)$, and therefore $B\approx (a/a^*_B)\ln(2a/d)$, with $a^*_B$ being the Bohr radius. Thus, the critical condition $B_c\sim 2$ is achieved when $a\gtrsim a_B$, which is compatible with the restriction $ak_F<1$ supposed in our calculations.

Direct comparison with the experiment is hindered because actually the length of the constriction is comparable or even larger than the wavelength, while the present calculations are strongly restricted by the requirement $ak_F\ll 1$. Calculations carried out within \textbf{the} 1D model system with specific potential $v(x)\propto \cosh^{-1}(x/a)$ for small but finite $ak_F$ show that the metastable state becomes polarized with increasing the length $a$~\cite{Sablikov}. The fact that the polarized state can be metastable is interesting since in this case the existence of the polarization does not contradict to the Lieb-Mattis theorem~\cite{Lieb-Mattis}.    

Nevertheless, simple qualitative arguments~\cite{Sablikov} show that the metastable state can manifest itself in a decrease of the conductance with the temperature if the metastable state is polarized. The effect occurs because the conductance in the polarized state is less than in the unpolarized state. This mechanism could explain the temperature dependence of the conductance observed experimentally when the $0.7\times2e^2/h$ anomaly is formed.

The existence of a metastable state in quantum point contacts was also seen in the numerical calculations of the conductance within the density functional approach for zero temperature~\cite{Starikov}. However, this state has not been identified and the grand potential of the system has not been investigated.

\acknowledgments
This work was partially supported by Russian Foundation for Basic Research (project No~11-02-00337) and programs of Russian Academy of Sciences.

\appendix

\section{Branching of the solutions in 1D model\label{appendix_branching_1D}}
Here we show that Eq.~(\ref{y(B)}) has two more roots, $y_s(B)$ and $y_{\bar s}(B)$, in addition to $y_0(B)$. They are branched from the function $y_0(B)$. Let us seek the solutions of Eq.~(\ref{y(B)}) in the vicinity of the function $y_0(B)$ by representing the sought function $y(B)$ in the form
\begin{equation}
y(B)=y_0(B)+\delta y(B)\,.
\end{equation}
If $\delta y\ll y_0$, the function $F(y,B)$ can be expanded in powers of $\delta y$
\begin{equation}
 F(y,B)=y_0+F_{y}'\delta y+\frac{1}{2}F_{yy}''\delta y+\frac{1}{6}F_{yyy}'''\delta y+\dots\,.
\end{equation} 
As a result, Eq.~(\ref{y(B)}) is transformed to
\begin{multline}
 \delta y\left\{1-F_{y}'^2-\frac{1}{2}F_{y}'F_{yy}''(1+F_{y}')\delta y\right.\\
\left.-\frac{1}{6}F_{y}'\left[F_{yyy}'''(1+F_{y}'^2)+3F_{yy}''\right]\delta y^2-\dots\right\}=0.
\end{multline}

The trivial solution of this equation, $\delta y=0$, corresponds to the root $y_0(B)$. If $\delta y\ne 0$, additional infinitely small solutions can exist when $1-F_y'^2$ is arbitrary small. Hence the point where $F_y'^2=1$ is a branching point of $y(B)$. Since according to Eq.~(\ref{F(yB)}) $F_y'<0$, the necessary condition for the branching point to exist is:
\begin{equation}
F_y'|_{y=y_0(B)}=-1.
\label{append_1}
\end{equation}
This is also an equation determining the critical value of the interaction parameter $B=B_c$ at which the function $y(B)$ has the branching point. In this point $y=y_c\equiv y_0(B_c)$.

By expanding $F(y,B)$ in two variables in the vicinity of the critical point ($y=y_c, B=B_c$) one can show that: 

(i) Eq.~(\ref{append_1}) has always one solution, 

(ii) near the critical point, the function $y(B)$ has the form
\begin{equation}
 y-y_c\simeq \pm C(B_c)(B-B_c)^{1/2},
\label{A_ys_B}
\end{equation}
where $C(B_c)$ is expressed in terms of derivatives of $F(y,B)$ at the critical point:
\begin{equation}
  C(B_c)=6\frac{-2F_{yB}''-F_{yy}''F_{B}'}{3F_{yy}''^2+2F_{yyy}'''}\Biggr|_{y=y_c, B=B_c}.
\end{equation}

\section{Stability of the solutions\label{appendix_stability}}
In the self-consistent field approach~\cite{Mermin}, the grand potential is minimized over the trial density matrices, which are chosen in the form of the equilibrium density matrix for non-interacting particles with an effective Hamiltonian of the form
\begin{equation}
 \hat{H}_{eff}=\sum\limits_{\mathbf{k},s,\mathbf{k'},s'}\gamma_{\mathbf{k},s,\mathbf{k'},s'} \hat{c}^{\dag}_{\mathbf{k},s} \hat{c}_{\mathbf{k'},s'}\,,
\end{equation}
where $\hat{c}^{\dag}_{\mathbf{k},s}$ and $\hat{c}_{\mathbf{k},s}$ are fermion creation and annihilation operators, $\gamma_{\mathbf{k},s,\mathbf{k'},s'}$ is a Hermitian matrix that is to be determined by the way of minimizing the grand potential $\Omega$ with respect to $\gamma_{\mathbf{k},s,\mathbf{k'},s'}$. In the calculations, it is convenient to use the matrix $\rho$ associated with the matrix $\gamma$ as follows:
\begin{equation}
 \rho_{\mathbf{k},s,\mathbf{k'},s'}=1/(1+\exp[\gamma_{\mathbf{k},s,\mathbf{k'},s'}])\,.
\end{equation} 
The requirement that the first variation of $\Omega$ is zero leads to the equation for single-particle wave functions. The stability of these states is investigated by analyzing the second variation $\delta^2\Omega$.

First we assume that the system is finite. $\delta^2\Omega$ can be written as
\begin{equation}
 \delta^2\Omega=\frac{1}{2}\!\sum\limits_{\substack{\mathbf{k},s,\mathbf{p},\sigma\\\mathbf{k'}\!,s'\!,\mathbf{p'}\!,\sigma'}}\! \delta\rho_{\mathbf{p}\sigma,\mathbf{k}s}\langle \mathbf{k}s,\mathbf{p}\sigma|X|\mathbf{p'}\sigma',\mathbf{k'}s'\rangle \delta\rho_{\mathbf{p'}\sigma',\mathbf{k'}s'},
\end{equation}
where $\delta\rho_{\mathbf{p}\sigma,\mathbf{k}s}$ is the variation of the matrix $\rho$ and the matrix $X$ is defined as
\begin{equation}
\begin{split}
\langle  &\mathbf{k}s,\mathbf{p}\sigma|X|\mathbf{p'}\sigma',\mathbf{k'}s'\rangle=\frac{\varepsilon_k-\varepsilon_p}{\rho_p^F-\rho_k^F}\,\delta_{\mathbf{k}\mathbf{p'}} \delta_{\mathbf{p}\mathbf{k'}} \delta_{s\sigma'}\delta_{s'\sigma}\\
&+\langle \mathbf{k}s,\mathbf{k'}s'|V_{ee}|\mathbf{p}\sigma,\mathbf{p'}\sigma'\rangle - \langle \mathbf{k}s,\mathbf{k'}s'|V_{ee}|\mathbf{p'}\sigma',\mathbf{p}\sigma\rangle.
\end{split}
\end{equation} 

The second variation $\delta^2\Omega$ is positive if the matrix $X$ is ​​positive definite. Since $X$ is a Hermitian matrix, it is positive defined if and only if the eigenvalues ​​of $X$ are positive. This means that the equation for eigenvectors
\begin{equation}
 \sum\limits_{\mathbf{k'},s', \mathbf{p'},\sigma'}\!\langle \mathbf{k}s,\mathbf{p}\sigma|X|\mathbf{p'}\sigma',\mathbf{k'}s'\rangle \alpha_{\mathbf{p'}\sigma',\mathbf{k'}s'}=\lambda_{\mathbf{k}s,\mathbf{p}\sigma}\alpha_{\mathbf{k}s,\mathbf{p}\sigma}
\end{equation} 
has positive eigenvalues $\lambda_{\mathbf{k}s,\mathbf{p}\sigma}>0$.

In the case of the 1D model, the matrix $X$ has the form:
\begin{equation}
\begin{split}
\langle ks,p\sigma &|X|p'\sigma',k's'\rangle=\frac{\varepsilon_k-\varepsilon_p}{\rho_p^F-\rho_k^F}\delta_{kp'}\delta_{pk'}\delta_{s\sigma'}\delta_{s'\sigma}\\&+\bar{v}aA_{ks}^*A_{k's'}^*A_{p\sigma}A_{p\sigma}\left(\delta_{s\sigma}\delta_{s'\sigma}-\delta_{s\sigma'}\delta_{s'\sigma}\right),
\end{split}
\label{X_L}
\end{equation} 
where $A_{ks}$ is the amplitude of the wave function at $x=0$. According to Eq.~(\ref{psi_1D})
\begin{equation}
 A_{ks}=\sqrt{\frac{2}{L}}\frac{1}{\sqrt{1+Q_{\bar s}^2/k^2}}\,.
\end{equation} 
Since $A_{ks}$ decreases as $L^{-1/2}$ with increasing $L$, the second term in Eq.~(\ref{X_L}) vanishes in the limit $L\to \infty$ and the matrix $X$ becomes diagonal. The eigenvalues of $X$ are positive since $(\varepsilon_k-\varepsilon_p)/(\rho_p^F-\rho_k^F)>0$.

This conclusion is straightforwardly generalized to the systems of higher dimensionality.


\begin{thebibliography}{20}
\bibitem{note1}
The case where $A+B<0$ needs a separate consideration since in this case the effective potential of the interaction region becomes negative and the bound state arises, which is not included in Eq.~(\ref{psi_1D}). The emergence and role of the localized states is considered below in Sec.~\ref{1Dsystem}.
\end{thebibliography}
\end{document}